\documentclass[preprint,showpacs,preprintnumbers]{revtex4}
\usepackage{graphicx}
\usepackage{dcolumn}
\usepackage{bm}
\usepackage{amsmath}
\usepackage{amsfonts}
\usepackage{amssymb}

\input{tcilatex}

\begin{document}

\title{Purcell factor enhanced scattering efficiency in optical microcavities%
}
\author{T.J. Kippenberg$^{1}$, A.L. Tchebotareva$^{2}$, J. Kalkman$^{2}$, A.
Polman$^{2}$, K.J. Vahala$^{1}$ }
\email{polman@amolf.nl,vahala@its.caltech.edu,}
\affiliation{$^{1}$ Thomas J. Watson Laboratory of Applied Physics,\ California Institute
of Technology,\ Pasadena, CA 91125.}
\affiliation{$^{2}$ Center for Nanophotonics, FOM Institute AMOLF, Kruislaan 407, 1098SJ,
Amsterdam, The Netherlands.}

\begin{abstract}
Scattering processes in an optical microcavity are investigated for the case
of silicon nanocrystals embedded in an ultra-high Q toroid microcavity.
Using \ a novel measurement technique based on the observable
mode-splitting, we demonstrate that light scattering is highly preferential:
more than 99.8\% of the scattered photon flux is scattered into the original
doubly-degenerate cavity modes. The large capture efficiency is attributed
to an increased scattering rate into the cavity mode, due to the enhancement
of the optical density of states over the free space value and has the same
origin as the Purcell effect in spontaneous emission. The experimentally
determined Purcell factor amounts to 883. 
\end{abstract}

\maketitle

Optical microcavities confine light both temporally and spatially and find
application in a variety of applied and fundamental studies, such as
photonics, cavity quantum electrodynamics, nonlinear optics and sensing\cite%
{VahalaNature}. In nearly all embodiments of microcavities, such as
microdisks, microspheres, micropillars or photonic crystals, sub-wavelength
defect centers are present, either caused by intrinsic material
irregularities, fabrication imperfection or intentionally induced (such as
quantum dots). The concomitant refractive index perturbations lead to
scattering, which increases the cavity loss rate. In this manuscript we
analyze the effect of scattering in an optical microcavity, using a toroid
microcavity containing silicon nanocrystals (Si NCs) as scattering centers.
Using a novel measurement technique we demonstrate for that light scattering
is highly preferential; $99.8\%$ of all scattered light is scattered into
the original eigenmodes. This value cannot be explained by the existing
geometrical optics theory\cite{Gorodetsky}. A novel theoretical analysis
shows that the observed enhanced scattering rate into the original cavity
mode is due to the enhancement of the optical density of states (DOS) over
the free space value, and therefore has the same origin as the
Purcell-effect in spontaneous emission. The presented experimental and
theoretical results establish the significance of the Purcell-factor for
scattering processes within a microcavity and constitute the highest
experimentally measured Purcell factor to date ($883$).

It is a well known phenomenon \cite{Gorodetsky,Weiss,Kippenberg} that the
resonances of whispering gallery mode (WGM) microcavities such as droplets,
microspheres, microtoroids or microdisks appear as doublets. The doublet
splitting is due to lifting of the two-fold degeneracy of the clockwise (CW)
and counter-clockwise (CCW) whispering-gallery-modes (WGMs) that occurs when
these modes are coupled due to scattering, either by intrinsic or surface
inhomogeneities. In the Rayleigh limit (particle radius $r\ll\lambda$), this
leads a scattered power $P$ per unit solid angle $\Omega$ as given by $\frac{%
dP}{d\Omega}=\frac{1}{4\pi\epsilon}k^{4}\left\vert \vec{p}\right\vert
^{2}\cos^{2}(\theta)$, with $\epsilon$ the dielectric constant of the
medium, $k$ the magnitude of the wave vector, and $\theta$ the scattering
angle. As discussed in reference \cite{Gorodetsky} this leads to doublet
splitting due to lifting of the CW and CCW mode degeneracy, since part of
the scattered light is channeled back into the original pair of eigenmodes,
leading to an observable mode splitting. In what follows, it is analyzed how
the observable mode spitting can be used to infer the capture efficiency ($%
\eta$), which is defined as the fraction of light scattered into the
original eigenmodes (and which, therefore, does not contribute to cavity
losses). For a single nanoparticle scattering cross section $\sigma_{scat}$,
and a number density $N$, the total scattering rate is given by $%
\gamma_{tot}^{-1}=\sigma _{scat}N\frac{c}{n}$ where $c$ is the speed of
light. The mode splitting (in the frequency domain) is then given by 
\begin{equation}
\gamma^{-1}=\frac{1}{2}\eta\gamma_{tot}^{-1}
\end{equation}

The factor of ${\frac{1}{2}}$ takes into account that the scattering of
light into original eigenmodes (i.e., self-coupling) does not contribute to
the observed mode splitting. Owing to the small size of the scattering
centers in comparison to the wavelength of light, it is assumed that
scattering is equally divided into CW and CCW direction. The dissipation
rate of light not scattered into the cavity mode is given by: 
\begin{equation}
\tau ^{-1}=(1-\eta )\gamma _{tot}^{-1}
\end{equation}%
In addition, $1/\tau _{0}$ will describe losses associated with absorption,
and the cumulative effect of these processes $(1/\tau _{0}+1/\tau )$ causes
a reduction in cavity $Q$. The degree to which the scattering process
couples the initially degenerate cavity modes can be described by the modal
coupling parameter $\Gamma $ \cite{Kippenberg}, which can be expressed by:

\begin{equation}
\Gamma=\left( \frac{\frac{1}{2}\eta\gamma_{tot}^{-1}}{\tau_{0}^{-1}+(1-\eta)%
\gamma_{tot}^{-1}}\right)  \label{Equation_Gamma}
\end{equation}
$\Gamma$ thus reflects the relative \textquotedblleft
visibility\textquotedblright\ of the doublet as appearing in the
under-coupled resonance spectrum; measurement of this parameter is described
in detail in refs. \cite{Kippenberg,Weiss}.

The capture efficiency $\eta $ can be retrieved from $\Gamma $ as follows.
In the presence of other loss channels, such as residual material
absorption, two limits of Eq. \ref{Equation_Gamma} can be considered. First,
if the cavity losses are dominated by absorption,i.e. $\tau _{0}^{-1}\gg
(1-\eta )\gamma _{tot}^{-1}$, Eq. \ref{Equation_Gamma} simplifies to:%
\begin{equation}
\Gamma =\frac{\eta \gamma _{tot}^{-1}}{2\tau _{0}^{-1}}
\label{Equation_AbsorptionLimited}
\end{equation}%
i.e. the doublet visibility, measured for different cavity modes, increases
linearly with intrinsic Q (if scattering is constant for all modes). In this
regime, a \textit{lower bound }of the capture efficiency $\eta $ can be
found, given by\ $\eta >\frac{2\gamma _{tot}\Gamma }{\tau _{0}}$. Second, in
the scattering-limited case, for which $\tau _{0}^{-1}\ll (1-\eta )\gamma
_{tot}^{-1}$ it follows from Eq. \ref{Equation_Gamma} that: 
\begin{equation}
\eta =\frac{2\Gamma }{1+2\Gamma }  \label{Equation_Efficiency}
\end{equation}%
In this regime an improved \textit{lower bound} of the capture efficiency $%
\eta $ can be inferred from measurements of $\Gamma $ (with the accuracy
given by the amount residual absorption). Significantly, in the case where
the residual absorption (i.e. $\tau _{0}$) is known and is identical for all
cavity modes (i.e. an intrinsic property of the resonator) $\eta $ can be
inferred exactly. It can then be determined by measuring $\Gamma $ for modes
with various amounts of scattering ($\gamma _{tot}$), and inferring $\eta $
via the functional relation of $\Gamma (Q,\gamma _{tot})$ as given by Eq. %
\ref{Equation_Gamma}.

To test the above model, we have investigated the scattering processes of SiO%
$_{2}$ toroid microcavities. These microcavities exhibit ultra-high-Q
whispering-gallery type modes \cite{Armani}, and can be used as ultralow
threshold chip based Raman lasers, optical parametric oscillators, or erbium
microlasers. Details on fabrication and on the coupling technique (using
tapered optical fibers) can be found in Refs.\cite{Armani,Spillane}

Figure 1 shows measurement of the $\Gamma $-parameter for an undoped \ SiO$%
_{2}$ 50-$\mu $m-diameter toroid microcavity, measured for successive
fundamental resonances with different Q in the 1550 nm band. A cavity
resonance scan is shown in the upper right panel of Fig. 1, and shows the
typical doublet splitting of \symbol{126}10 MHz observed for pure SiO$_{2}$
toroids. The data in the main panel of Fig. 1 clearly follow a linear
relationship, indicating that the cavity resonances follow
absorption-limited behavior (attributed to adsorbed water and OH on the
surface of the toroid\cite{RokhsariAPL}). The scattering rate $\gamma
^{-1}/2\pi $ derived from the data is plotted in the lower right figure (as $%
Q_{split}=\omega \gamma $), and is indeed to a very good approximation the
same for all modes. From the highest observed doublet splitting ($\Gamma =28$%
) the lower estimate of the capture efficiency is $\eta >96.4\%$. This value
cannot be explained by the quasi-geometrical estimations of Ref. \cite%
{Gorodetsky} which predict a \textit{maximum} capture efficiency of $90\%$.
This model assumes that scattered light obeys a Rayleigh-type angular
distribution, and can couple back into the CW and CCW modes, provided the
scattering angle $\theta $ is within the critical angle $\phi $ of the mode
i.e. $\theta $$<$$\phi $. This model, while adequate to describe losses of a
waveguide-bend, is (as will be shown below), incomplete as it neglects the
periodic nature of light scattering in a microcavity.

To infer the capture efficiency more exactly, measurements were performed in
the scattering-limited regime, by fabricating SiO$_{2}$ toroid
micro-cavities doped with Si NCs. Si NCs exhibit quantum-confined
photoluminescence (PL) in the visible and near-infrared, and have various
potential applications in photonic and electronic devices\cite{Polman}. Si
NCs do not posses significant absorption transitions at $\lambda =1.5$ $\mu
m $ and have a high refractive index relative to the SiO$_{2}$ matrix ($%
n=3.48$ vs. $n=1.44$) and thus act (in the 1550-nm band) as strong
scattering centers. The Si NC doped cavities were made by ion-implantation
of $900$ keV Si$+$ ions (fluence $9.1\times 10^{16}$ $cm^{-2}$) into a
thermally oxidized Si wafer (2 $\mu m$ oxide), followed by annealing\cite%
{Min} and toroid fabrication. In order to confirm the presence of Si NCs
after fabrication, 2-D cross-sectional PL images were measured using a
confocal imaging microscope (using a rhodamine doped index matching oil).
Figure 2(a) shows a cross sectional image of the integrated PL in the $%
650-690$ nm band, taken in the toroid's equatorial plane. A bright
luminescent ring is observed, characteristic of quantum-confined emission
from Si NCs embedded in the toroid. The emission spectrum (cf. Fig.2) peaks
at $\lambda =675$ nm, corresponding to a NC diameter of $\sim $3 nm. A cross
sectional image of the toroid's minor ring is shown in Fig. 2 ( resolution
of $400\times 850$ nm). Clear NC PL is again observed inside the toroidal
ring. The outer PL ring in Fig. 2 corresponds to emission from the rhodamine
dye adsorbed onto the surface, and serves to determine the cavities' outer
contour. We find that while the NC PL is inhomogeneously distributed, Si NC
PL is observed throughout the entire toroidal volume.

The optical resonances of Si NC doped microcavities exhibited splitting
frequencies as large as 1 GHz, nearly two orders of magnitude larger than
for the undoped SiO$_{2}$ toroids. This confirms that scattering centers
(here Si NCs) are responsible for the observed strong modal coupling (i.e. $%
\Gamma\gg 1$). The highest observed modal coupling parameter was $\Gamma=50$%
. Correspondingly, according to Eq. \ref{Equation_Efficiency}, the capture
efficiency is $\eta>98\%$. Thus despite strong scattering from the NCs, long
photon storage times are still achievable. Indeed, Q-factors $>10^{7}$ are
observed for most measured resonances.

In order to obtain an even more exact value of the capture efficiency, two
different sets of transverse cavity modes (attributed to the radial mode
index $n=1$ and $n=2$) were characterized with progressing angular mode
numbers ($\ell ,\ell +1,\ell +2,...$). Due to the inhomogeneous distribution
of the NCs (cf. Fig. 2), these modes are dominated by differing levels of
scattering. However, due to the presence of water and OH adsorbed onto the
cavity surface, each set of radial modes experiences the same amount of
residual absorption. $\Gamma $ and $Q$ measurements for each of these
resonances are shown in Fig. 3 where the splitting frequency is expressed as
splitting quality factor $Q_{split}=\omega \gamma $. The solid line in Fig.
3 is a two-parameter ($\eta $ and $\tau _{0}$ ) fit of Eq. \ref%
{Equation_Gamma} applied to the high-Q experimental data (attributed to the $%
n=1$ radial modes), and excellent agreement is obtained for $\eta =99.42\%$ $%
(\pm 0.04\%)$ and $\tau _{0}=115$ ns ($\pm 3$ ns) (i.e.$Q_{0}=1.4\times
10^{8}$). The deviation between fit and data for the lower-Q data is
attributed to higher-order radial modes ($n=2$), which possess increased
intrinsic (OH and water) absorption losses, owing to their slower decaying
field amplitude outside the cavity. The mode identification is also
consistent with the inhomogeneous distribution of NCs, which causes
increased scattering for higher order radial modes (cf. Fig. 2). These modes
follow absorption-limited Q behavior, represented by a constant $Q_{scat}$,
(and a linear dependence of $\Gamma (Q)$, see inset Fig. 3). The inferred
intrinsic cavity lifetime $\tau _{0}$ is in good quantitative agreement with
recent estimates of the absorption loss due to water adsorbed onto the
cavity surface.\cite{RokhsariAPL} The remarkably high capture efficiency of $%
99.42\%$ found here clearly demonstrates that optical scattering in
microcavities occurs preferentially into cavity eigenmodes.

This observation can be understood by the following qualitative reasoning.
For a single nanoparticle coupled to a microcavity mode, the temporal
confinement of light implies periodic scattering with the temporal
periodicity given by the cavity round-trip time. Conversely, one can treat
the problem as scattering by a chain of $M$ equidistantly spaced scattering
centers (spatial periodicity of $L=2\pi R$). The incident optical field will
induce a dipole moment $\vec{p}$, which is in phase at each scattering site.
The number ($M$) of excited dipoles is approximately given by the cavity
lifetime divided by the round trip time ($T$) i.e.\qquad%
\begin{equation}
M\tilde{=}\frac{\tau c}{2\pi n_{eff}R}=\frac{Q\lambda}{4\pi^{2}Rn_{eff}}
\end{equation}

This problem can be recognized as 1-D scattering of a periodic lattice and
is well known in solid state physics or diffractive optics. The interference
of all $M$ scattering centers, leads to a scattering intensity (i.e.
diffraction pattern) with a width of the corresponding principal diffraction
maxima as given by: $\Delta\theta=\frac{2\pi}{M}$. Therefore, the on-axis
scattering rate is enhanced by the cavity by a factor of $M$, compared to
the peak scattering rate of a single emitter (dipole) in free space i.e. $%
\left( \frac{dP}{d\Omega}\right) _{\theta=0}^{(M)}=M\left( \frac{dP}{d\Omega 
}\right) _{\theta=0}^{(\sin gle)}$. This angular narrowing of the spatial
emission profile due to a collection of dipoles has been treated in the
context of super-radiance\cite{Rehler} and pertains both to an atomic
(quantum) dipole as well as to a classical system of radiators. Note that $M$
is proportional to $Q/V_{m}$ where $V_{m}$ is the mode volume. This
motivates a second interpretation, by considering the effect of DOS in the
scattering process.

If we consider $\vec{E}_{d}$ to be the dipole field, then the probability of
scattering into a designated plane wave with wave vector $\vec{k}$ and
polarization $\vec{\sigma}$ is proportional to the integral $\epsilon\int 
\vec{E}_{k,\sigma}\cdot\vec{E}_{d}dV$.\cite{Vahala} The total scattering
rate is obtained by summation over all possible modes. Correspondingly, if
the DOS is altered -- e.g. due to the presence of a cavity -- the emission
into certain modes is enhanced, while emission into others is suppressed.
The enhancement of scattering is therefore given by the ratio of the optical
DOS in the microcavity to the value of free space, which at resonance is
given by\cite{PURCELL} 
\begin{equation}
F=\frac{3}{4\pi^{2}}\frac{Q}{V}\left( \frac{\lambda}{n}\right) ^{3}
\end{equation}
i.e. the well known \textit{Purcell factor}. An explicit mathematical
derivation of the enhancement as outlined here has also been undertaken and
will be communicated elsewhere.

Note that an excited atom within a high-Q microcavity (in the regime of weak
coupling), will also preferentially emit into the cavity mode. This is
referred to as the Purcell\textit{\ effect}, and the corresponding
shortening of the spontaneous emission\ (SE) lifetime is given by the
Purcell factor as first proposed in\ Ref. \cite{PURCELL}. Note that the
origin of preferential emission for an atomic and a classical dipole are
identical; for an atom the enhancement of SE follows from classical
electrodynamics, and results from the enhanced DOS of the vacuum field\cite%
{MilonniBook}. Furthermore, atomic SE enhancement can be interpreted in
terms of a chain of radiating dipoles (as done here in the case of
scattering) when using the image method, which replaces the cavity mirrors
(in case of a Fabry- P\'{e}rot cavity) by an infinite chain of image dipoles%
\cite{Milonni}.

Having established the relation of preferential scattering and the optical
density of states, we can now relate the capture efficiency $\eta $ to the
Purcell factor via:%
\begin{equation}
\eta =\frac{F}{F+1}
\end{equation}%
as in the case of an atomic emitter ($\beta $ factor). The experimentally
found value of $\eta =99.42$ $\pm 0.04\%$ (cf. Fig. 3) then corresponds to a
Purcell factor $F=172\pm 10$. This represents the highest measured value of
this parameter to date. Theoretically, the Purcell factor for the toroid
microcavities used in the experiments $(Q=1.3\times 10^{8},D=72\mu m$ , $%
A_{eff}=14\mu m^{2})$ equals $F=2060$, corresponding to $\eta =99.95\%.$ The
discrepancy between this and the experimentally found value is likely to be
caused by the presence of minute absorption (originating from defect-related
mid-gap states, or from two-photon absorption in the Si NCs). Taking
absorption into account, the modal coupling parameter is given by:

\begin{equation}
\Gamma =\left( \frac{\frac{1}{2}\eta \gamma _{tot}^{-1}}{\tau
_{0}^{-1}+(1-\eta )\gamma _{tot}^{-1}+\frac{\sigma _{abs}}{\sigma _{scat}}%
\gamma _{tot}^{-1}}\right)  \label{GammaAbs}
\end{equation}%
with $\sigma _{abs}$ being the absorption cross section. Repeating the fit
of data in Fig 3. with Eqs. 9 we obtain, $\tau _{0}^{-1}=115\pm 3.5$ $ns,$ $%
\left( \frac{\sigma _{abs}}{\sigma _{scat}}\right) ^{-1}=211\pm 15$ and $%
\eta =99.89\pm 0.04\%.$The equivalent Purcell factor is $F=883\ (+484/-229)$
and is in better agreement with the theoretical prediction. The remaining
discrepancy is attributed to the fact that scattering centers located at
different positions in the toroid experience the \textit{local} density of
states, which is lower than the value given by eqs. 7. Therefore the
measured value corresponds to the local D.O.S averaged over the spatial
distribution of the scattering centers.

In conclusion, we have analyzed and determined for the first time the
capture efficiency of scattered light in an optical micro-cavity. For a
toroidal microcavity doped with silicon nanocrystals, $99.89\%$ of the
scattered light is preferentially scattered into a cavity mode, equivalent
to a more than two order of magnitude increase of the scattering rate into
the microcavity mode with respect to the free space scattering rate. This
enhancement is found to be related to the ratio of the optical density of
states in the microcavity to that of free space, as given by the
Purcell-factor, and is an intrinsic property of any\textit{\ }microcavity.
Equally important, our results demonstrate that the Purcell factor of an
whispering-gallery mode microcavity can be measured directly via the
observed modal coupling.

\subsection{\textbf{Acknowledgements}}

We thank Prof. S. Roorda (Universit\'{e} de Montr\'{e}al) for Si ion
implantation. This work was funded by the DARPA, the NSF, and the Caltech
Lee Center. The Dutch part of this work is part of the research program of
FOM, which is financially supported by NWO. A.T. is grateful to the Fonds
NATEQ (Qu\'{e}bec, Canada) for a postdoctoral scholarship. T.J.K.
acknowledges a postdoctoral scholarship from the IST-CPI. The authors
acknowledge Dr. Oskar Painter for valuable discussions.

\bigskip 
\bibliographystyle{apsrev}
\bibliography{ReferencesScatteringPurcellEffect}

\newpage

\begin{figure}[tbp]
\begin{center}
\includegraphics[
width=3.039in
]{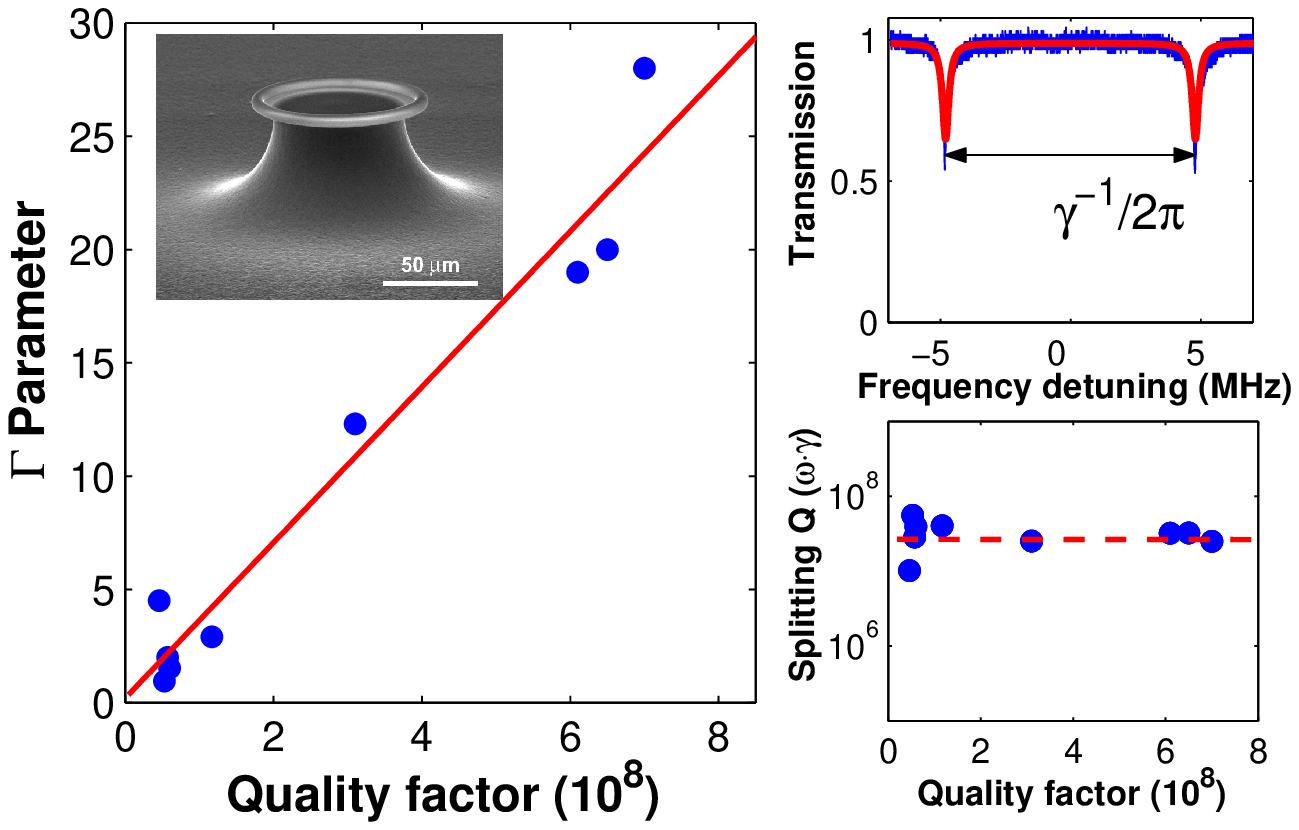}
\end{center}
\caption{Left figure: Modal coupling parameter (or doublet \textquotedblleft
visibility\textquotedblright) $\Gamma$ as a function of Q for several
fundamental modes of a SiO$_{2}$ toroid microcavity. A SEM micrograph of a
toroid microcavity is shown as inset. Upper right: characteristic spectral
scan, showing a typical mode splitting of $\sim$10 MHz and $\Gamma\sim$30.
Lower right panel: Splitting $Q_{split}$ $(=\protect\omega\protect\gamma)$
as a function of $Q$, which is nearly identical for all modes of the
resonator.}
\end{figure}

\begin{figure}[tbp]
\begin{center}
\includegraphics[
width=2.7294in
]{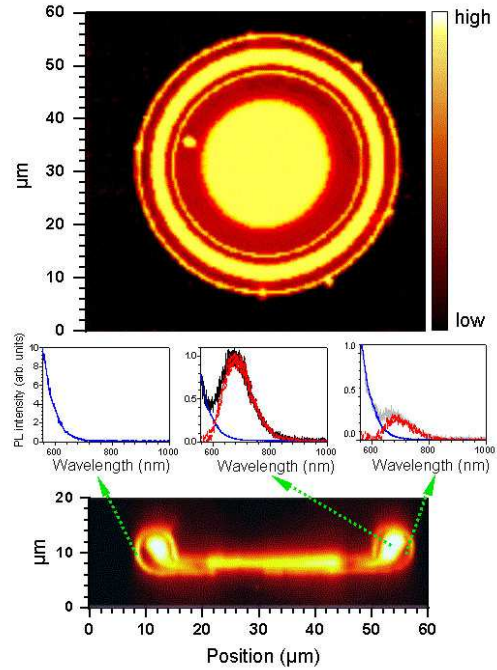}
\end{center}
\caption{Cross-sectional confocal PL images taken on a 72-$\protect\mu m$%
-diameter Si NC doped SiO$_{2}$ toroidal microcavity.Photoluminescence is
collected in 650-690 nm band. (a) x-y coss section. (b) y-z cross section.
Both images are taken with the toroid immersed into index-matching oil doped
with rhodamine. The insets show PL spectra taken at characteristic locations
in the toroid. The outer bright line in both images is attributed to the PL
of the rhodamine dye and serves to determine the cavity's outer contour.}
\end{figure}

\begin{figure}[tbp]
\begin{center}
\includegraphics[
width=2.8374in
]{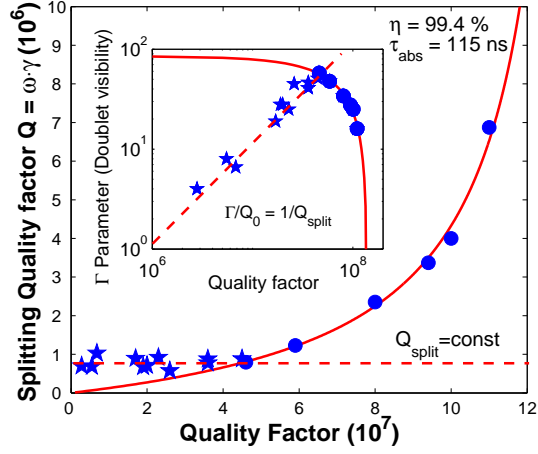}
\end{center}
\caption{Splitting quality factor ($Q_{split}=\protect\omega \protect\gamma $%
) as a function of $Q$ for the microcavity from Fig. 2. Inset: Same data,
plotted as $\Gamma $ as a function of $Q$. The solid line in both graphs is
a two-parameter fit using Eq. (\protect\ref{Equation_Gamma}) to the
scattering-limited WGMs (closed circles), yielding $\protect\tau %
_{_{0}}=115.2\pm 3$ ns, and $\protect\eta =99.42\pm 0.04\%$. The low-Q data
(stars), fitted with the dashed lines, correspond to higher-order
(absorption limited) radial modes (Eq. (\protect\ref%
{Equation_AbsorptionLimited})) }
\end{figure}

\end{document}